\documentclass[preprint2]{aastex63}

\received{June 1, 2019}
\revised{January 10, 2019}
\accepted{\today}

\submitjournal{ApJ}

\shorttitle{Low-Frequency Survey}
\shortauthors{Tremblay, C.D. et.al.}

\graphicspath{{./}{figures/}}

\begin{document}

\title{Nitric Oxide and other molecules: Molecular Modelling and Low Frequency Exploration using the Murchison Widefield Array}

\correspondingauthor{Chenoa Tremblay}
\email{chenoa.tremblay@csiro.au}

\author[0000-0002-4409-3515]{Tremblay, C.D.}
\affiliation{CSIRO Astronomy and Space Science, PO Box 1130, Bentley WA 6102, Australia}
\author[00000-0002-2094-846X]{Gray, M.D.}
\affiliation{Jodrell Bank Centre for Astrophysics, School of Physics and Astronomy, University of Manchester, Manchester M13 9PL, UK}
\affiliation{National Astronomical Research Institute of Thailand, 260 Moo 4, T. Donkaew, A. Maerim, Chiangmai 50180, Thailand}
\author[0000-0002-5119-4808]{Hurley-Walker, N.}
\affiliation{International Centre for Radio Astronomy Research, Curtin University, GPO Box U1987, Perth WA 6845, Australia}
\author[0000-0002-2670-188X]{Green, J.A.}
\affiliation{CSIRO Astronomy and Space Science, PO Box 1130, Bentley WA 6102, Australia}
\author[0000-0003-0235-3347]{Dawson, J.R.}
\affiliation{Department of Physics and Astronomy and MQ Research
Centre in Astronomy, Astrophysics and Astrophotonics, Macquarie University, NSW 2109, Australia}
\author[0000-0002-6300-7459]{Dickey, J. M.}
\affiliation{School of Maths and Physics,University of Tasmania, Hobart, TAS 7001, Australia}
\author[0000-0001-9429-9135]{Jones, P. A.}
\affiliation{Department of Physics and Astronomy and MQ Research
Centre in Astronomy, Astrophysics and Astrophotonics, Macquarie University, NSW 2109, Australia}
\author[0000-0002-8195-7562]{Tingay, S.J.}
\affiliation{International Centre for Radio Astronomy Research, Curtin University, GPO Box U1987, Perth WA 6845, Australia}
\author[0000-0003-4264-3509]{Wong, O.I.}
\affiliation{CSIRO Astronomy and Space Science, PO Box 1130, Bentley WA 6102, Australia}
\affiliation{International Centre for Radio Astronomy Research, UWA, 35 Stirling Hwy, Crawley, WA 6009, Australia}

\begin{abstract}
We present new molecular modelling for  $^{14}$NO and $^{15}$NO and a deep, blind molecular line survey at low radio frequencies (99--129\,MHz).  This survey is the third in a series completed with the Murchison Widefield Array (MWA), but in comparison with the previous surveys, uses four times more data (17\,hours vs. 4\,hours) and is three times better in angular resolution (1$'$ vs. 3$'$).  The new molecular modelling for nitric oxide and its main isotopologue has seven transitions within the MWA frequency band (although we also present the higher frequency transitions). Although we did not detect any new molecular lines at a limit of 0.21\,Jy\,beam$^{-1}$, this work is an important step in understanding the data processing challenges for the future Square Kilometre Array (SKA) and places solid limits on what is expected in the future of low-frequency surveys. The modelling can be utilised for future searches of nitric oxide.

\end{abstract}

\keywords{astrochemistry $-$ molecular data $-$ radio lines: stars: surveys $-$ ISM:molecules}

\section{INTRODUCTION }
\label{sec:intro}
Molecules are valuable tracers within our Galaxy to explore the chemical and physical environments of stars, dust, and gas. Published studies of molecular lines at low frequencies ($<$700\,MHz) are rare (e.g. \citealt{Marthi_OH55MHz}) and have included the contributions made with the Murchison Widefield Array (MWA;\citealt{Tremblay_PhD}). The science goals of these low frequency detections include: understanding the physical properties of the regions in which they are found, in particular those of high mass star formation; and understanding the formation mechanisms of molecules, such as sulfur and nitrogen bearing molecules and amino acids (as the building blocks of life). The goal of this project is to both understand the molecules that may be present and push the surveys to deeper levels than previous work.

The understanding of high-mass ($>$8-10\,M$_{\odot}$) star formation, which is still not well understood, can be improved by observations at frequencies less than 1\,GHz \citep{Codella_2015}. Seeing the need to understand the evolution of complex organic molecules in high-mass stars, \cite{Coletta_2020} studied 39 H{\sc ii} regions with the IRAM 30\,m Telescope that are in various stages of development.  They found the largest number of detections in ultra-compact (size: 0.05 $<$ R $\leq$ 0.1\,pc; density: n $\geq$ 10$^{4}$\,cm$^{-3}$) H{\sc ii} regions.  However, the higher temperatures in young high-mass stars create confusion from molecules that produce intense emission or a large number of transitions within the microwave and infrared frequency range, making it difficult to find new rare molecules. The lower frequency part of the radio spectrum is far less crowded. In addition, as we approach the regime where the ratio of the level spacing to the thermal energy of the gas is very small (i.e. small h$\nu$/kT) it is possible that lines may be more readily inverted \citep[e.g.][]{elitzur1992}, and we may wish to seek for signs of maser emission.

Therefore, a combination of surveys at the metre, centimetre, and millimetre wavelengths of the same regions of the sky, may unravel the mysteries of the formation mechanisms of high-mass stars through analysis of their chemical evolution.  As high-mass stars have a large impact on the initial mass function and chemical enrichment of a galaxy, obtaining a better understanding of their evolution within our Galaxy is important to galactic archaeology. With a combination of the low-frequency surveys with the MWA and new high-resolution surveys with the Australian Square Kilometre Array Pathfinder \citep{GASKAP}, we can start to probe the southern sky, with access to the inner Galactic Plane, in new detail.

From the molecular perspective, the motivation to observe at lower radio frequencies is not limited to complex molecules.  Rare simple molecules, such as the free radicals, CH ($\sim$724\,MHz), SH ($\sim$111\,MHz), or NO (107.4\,MHz - this work), are predicted to exhibit boosted emission similar to OH at 1.6\,GHz described more in Section 2) and were tentatively detected with our previous molecular line surveys with the MWA \citep{Tremblay_2017,Tremblay_2018}. Nitric Oxide (NO) is of particular interest because is expected to be widespread in the interstellar medium \citep{1995PhDT.........1M} and it plays an important role in the formation of hydroxylamine (H$_{3}$NO), which is an important molecule in the pathway of the formation of amino acids. Laboratory experiments have created enamines (precursors to amines) with deuterated formic acids in a neutral medium \citep{Himmele-1979}, suggesting that the search for formic acid and nitric oxide may point to the production of amines within circumstellar environments. NO is also of particular interest as it is thought to be critically important to primitive life on Earth \citep{Santana-2017}.

Despite this interest, and even with the precise frequencies of emission having been determined, little is known about the formation of NO, nor the primary emission frequencies we could expect to detect in interstellar space at metre wavelengths.  Both \cite{Quintana-Lacaci_NO_2013} and \cite{Chen_2014} summarise some theoretical modelling regarding the formation of NO in evolved stars.  They suggest that NO forms in shocked gas surrounding  Asymptotic Giant Branch (AGB) stars and Red Supergiants (RSGs) within the layer inhabited by OH masers (typically located at distances of 5--50\,R$_{\star}$).  They suggested a formation mechanism of OH + N $\rightarrow$ NO + H, which is a barrierless reaction. \cite{Cernicharo_2014} suggests a similar reaction for the formation of NO in the diffuse interstellar medium. However, contained in the astrochemical database KIDA (Kinetic Database for Astrochemistry; \citealt{KIDA}) are other possible formation routes of NO through cosmic ray interaction, photodissociation, and bimolecular reactions.

\begin{figure*}
\centering
\includegraphics[width=0.48\textwidth]{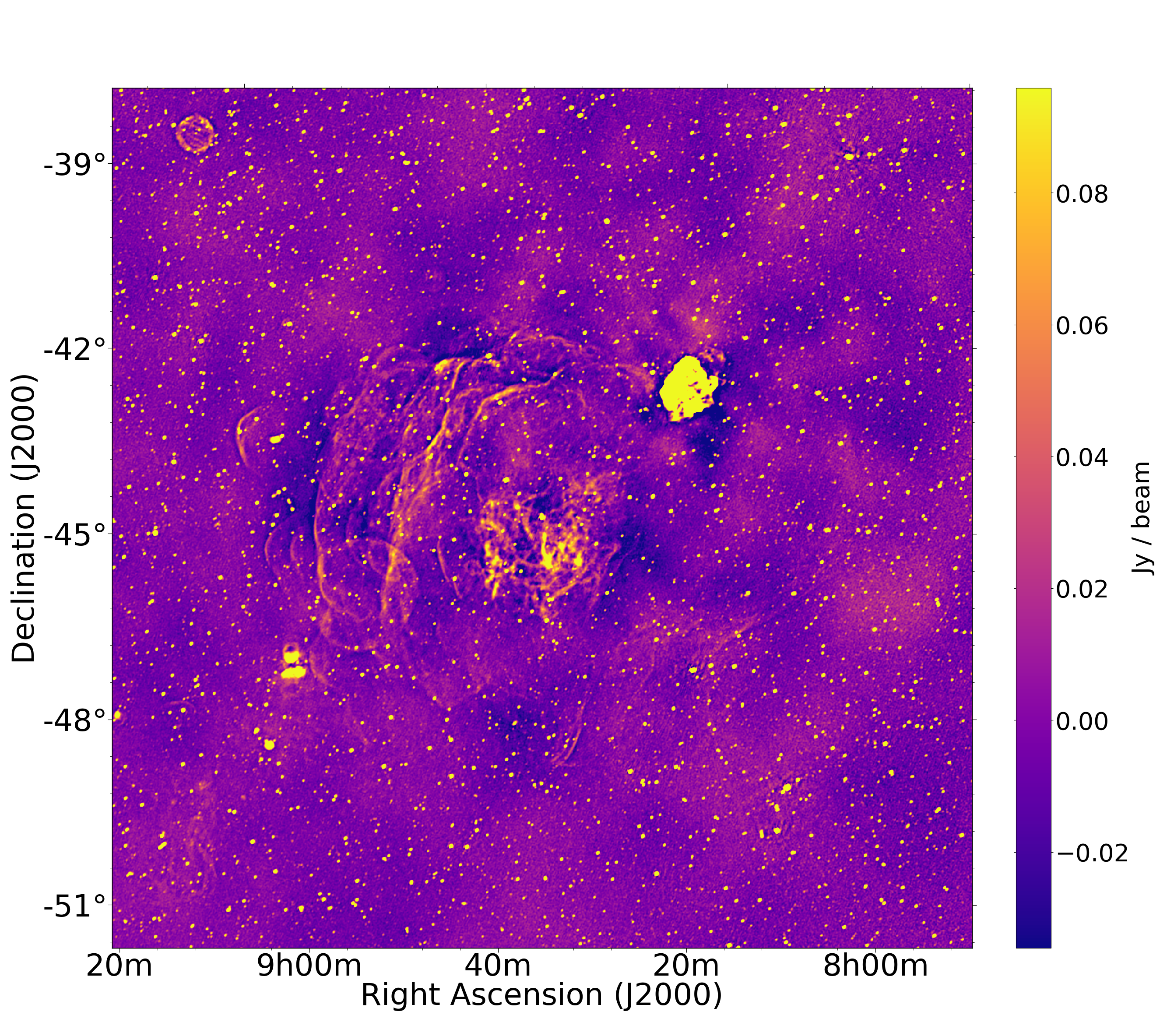}
\includegraphics[width=0.47\textwidth]{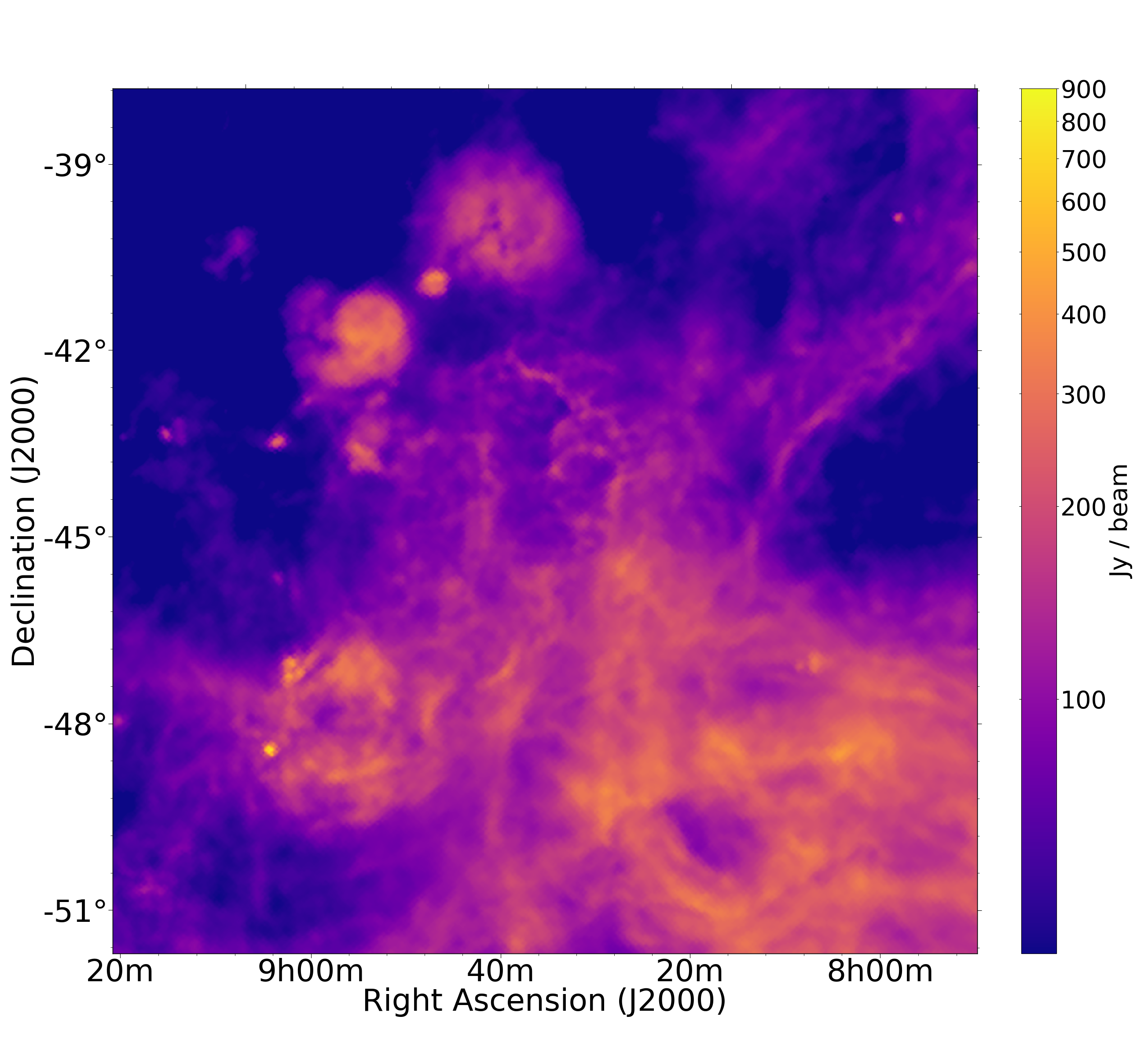}
\caption{Continuum image of the survey field, as observed with the Murchison Widefield Array, at a central frequency of 113.3\,MHz, with a bandwidth of $\sim$31\,MHz on the left and in optical H$\alpha$ on the right. The full-sky H$\alpha$ map (6' FWHM resolution) is a composite of the Virginia Tech Spectral line Survey (VTSS; \citealt{VTSS}) in the north and the Southern H-Alpha Sky Survey Atlas (SHASSA; \citealt{Finkbeiner_2003_Halpha}) in the south.  The molecular line survey covers $\sim$200\,square degrees centred on 08:35:20 --45:10:35 (J2000). }
\label{fig_Survey}
\end{figure*}

In our previous surveys with the MWA we observed the Galactic Centre \citep{Tremblay_2017} and the Orion Molecular Cloud Complex \citep{Tremblay_2018}. In this paper we focus on the region toward the Vela Constellation.  Previous to 1991 \citep{Murphy_1991}, the Vela constellation was thought to be devoid of the presence of spiral arms or spurs, with no star formation activity, however it has since become a region of intense scientific interest.  In the foreground is the wind swept H{\sc ii} region called the Gum Nebula, shown on the right-hand image in Fig. \ref{fig_Survey} as an $\sim$18\,degree \citep{Chanot_1983,Purcell_2015} bubble of intense H$\alpha$ emission about 200\,pc away. Then at an intermediate distance of about 400\,pc to 1\,kpc (depending on the filament) is the Vela Supernova Remnant high-velocity shock front (also known as the IRAS Vela Shell), shown in the left-hand image in radio continuum as a ``flower-like" structure \citep{Pakhomov_2012}.  Finally, in the background is the Vela Molecular Cloud Complex, which has been the subject of a large number of molecular line studies due to its unique molecular structures between 1-2\,kpc away. \cite{Yamaguchi_Gum_1999} studied the region in CO and found 82 molecular clouds and a number of cometary globules.  This has led us to study the region at low radio frequencies to see what we can unravel about its secrets.

In this paper we present new molecular modelling for NO transitions at frequencies between 3--4000\,MHz (Section 2), together with a blind molecular line survey of the 192 known molecular transitions in the band of 99--129\,MHz, over 200\,square degrees toward the Vela region (Sections 3,4,5). For the survey we use a total of 17 hours integration time, making it the deepest survey with the MWA to date. We then discuss the prospect for observations with Square Kilometre Array (SKA) Pathfinders and the SKA itself (Section 6).

\section{Nitric Oxide} 
Interstellar Nitric Oxide (NO; also known as nitrogen monoxide or the nitrosyl radical) was first observed by \cite{Liszt_1978} toward the star forming, and hydroxyl (OH) maser emitting, region of Sagittarius B2 using the 11\,m Kitt Peak radio telescope at 150.2 and 150.5\,GHz. Since then nitric oxide has been detected in thermal emission towards evolved stars \citep{Quintana-Lacaci_NO_2013,  Velilla_2015}, dark molecular clouds \citep{Gerin_1992}, the interstellar medium \citep{1995PhDT.........1M,Cernicharo_2014}, and star forming regions \citep{Ziurys_1991}. All of which are detected in the 100's of GHz frequencies.  The only tentative detection of low-frequency maser emission was with the MWA towards the Galactic centre \citep{Tremblay_2017} and Orion Molecular Cloud Complex \citep{Tremblay_2018}.

Based on work done by \cite{Meerts_1972} and \cite{Meerts_1976}, NO is one of the more prominent known molecules in the low-frequency range, being a highly reactive free radical. Despite little being known regarding which of the low-frequency molecular transitions are most likely to be detected in the interstellar medium, we would expect to see some at the frequencies detectable by the MWA. With this motivation, we present here new modelling of NO and its role in relation to known interstellar free radicals.

\subsection{New Molecular Modelling}
\begin{figure}
\includegraphics[width=0.47\textwidth]{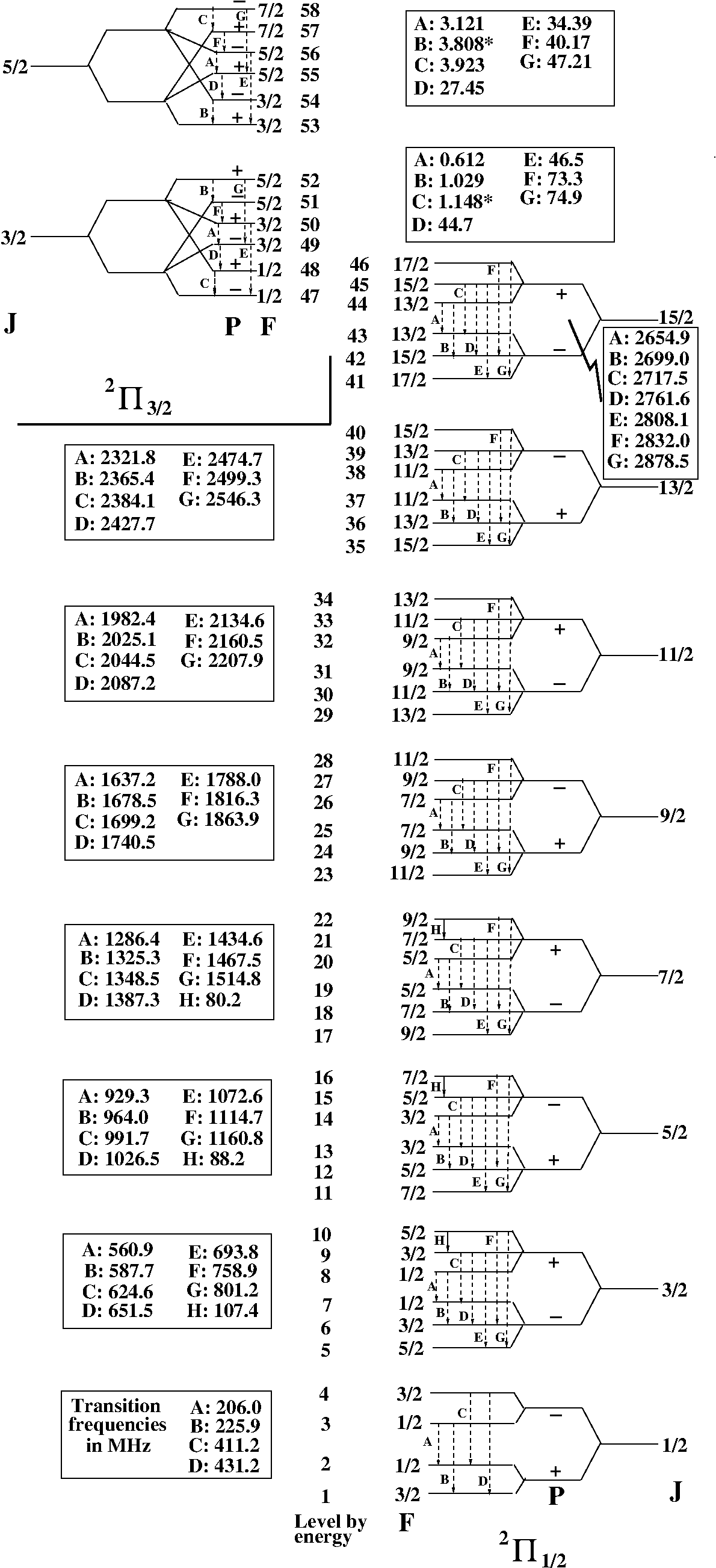}
\caption{The lowest 58 hyperfine energy levels of $^{14}$N$^{16}$O, showing the two rotational ladders resulting from spin-orbit coupling ($^2\Pi_{1/2}$ and $^2\Pi_{3/2}$), lambda doubling with components marked by the parity, or Kronig symmetry, $P$, and hyperfine structure, denoted by the quantum number $F$. Rotational levels (quantum number $J=\Omega + R$, where $R$ represents the nuclear framework rotation) are shown correctly ordered by energy, as are hyperfine levels within each group, but the rotational spacings, and all finer structure, are not to scale. }
\label{levels_14N16O}
\end{figure}

The similarity of the NO molecule to OH is due to the presence of a single unpaired electron, with one unit of orbital angular momentum and half a unit of spin, in both species. This results in two ladders of rotational levels, where the components of these angular momenta along the inter-nuclear axis are aligned ($^2\Pi_{3/2}$, or $\Omega=3/2$) and opposed ($^2\Pi_{1/2}$, or $\Omega=1/2$). The
lower-energy ends of these ladders, up to an energy of 133.5\,cm$^{-1}$ (or 192\,K), are shown in Fig.~\ref{levels_14N16O} together with 
details of lambda doubling and hyperfine structure.

The similarity to OH is closest for the isotopologue $^{15}$N$^{16}$O, in which the nuclear spin of the $^{15}$N is
1/2, as for the H-atom in OH. The resulting hyperfine structure has four levels per rotational level, with an F-quantum number that has integral values. The case of $^{14}$N$^{16}$O is more
complicated, because $^{14}$N has a nuclear spin of 1. Groups of six hyperfine levels usually result, labelled by a half-integral F-quantum number, as shown in 
Fig.~\ref{levels_14N16O}; only four hyperfine levels are possible in the rotational ground state. The energy-level data for Fig.~\ref{levels_14N16O} are drawn from the NO {\sc radex} file in the Leiden Atomic and Molecular Database \citep{2005A&A...432..369S} that in turn made use of data from the Cologne Database for Molecular Spectroscopy (CDMS;
\citealt{2005JMoSt.742..215M}). We note that the CDMS data use the Hund's case~(b) notation, instead of the case~(a) notation used in the present work. Levels with the case~(b) quantum number $N=J+1/2$ equate to the $^2\Pi_{1/2}$ (and $N=J-1/2$, to $^2\Pi_{3/2}$) for $J\leq 11/2$; the equivalence is reversed for higher $J$. The sign of the parity in the present work
agrees with the Leiden file, and appears to be reversed with respect to \citet{Meerts_1972}.

Fig.~\ref{levels_14N16O} includes transition frequencies in MHz for all hyperfine transitions that change parity. Additional transitions that do not change parity, but appear in the Leiden {\sc radex} file and have frequencies in the MWA range, are also shown. Transition data were cross-referenced from the Leiden file to data tables in \citet{Meerts_1972}.  Frequencies followed by an asterisk appear in neither the Leiden file nor the tables in \citet{Meerts_1972}, but were taken directly from the CDMS database.  Transitions of
NO within or near to the frequency range of the MWA are listed with more accurate frequencies and additional information in Table~\ref{hugetab}.

In summary, the intricate hyperfine and lambda doubling structure of NO, combined with its increased mass relative to OH and CH, leads to the presence of many allowed transitions with frequencies $<$1\,GHz that may be excited under typical interstellar and circumstellar temperatures. Many would be suitable candidates for low-frequency instruments, such as the MWA. Some NO transitions are, in fact, well below the ionospheric plasma frequency($\sim$30\,MHz). This new representation of the energy levels of NO will be used to motivate searches for the transitions identified, both with the MWA and at higher frequencies with SKA pathfinders, especially where simultaneous searches for NO, CH, and OH may be possible. 

\begin{table*}
\caption{Transitions of $^{14}$N$^{16}$O that are within the MWA frequency band, and are included in Fig.~\ref{levels_14N16O}, ordered here by increasing frequency.  The N$_u$ and N$_l$ refer to the energy order levels from Fig.~\ref{levels_14N16O}. }
\label{hugetab}
\resizebox{\textwidth}{!} {%
\begin{tabular}{lllrrrrc}
\hline
$\nu$        &         Upper Level                  &        Lower Level                 & N$_u$ & N$_l$ & T$_{up}$   & Einstein~A            & Notes   \\
MHz          & $^2\Pi_{\Omega}, J=j', F(P)$           & $^2\Pi_{\Omega}, J=j", F(P)$         &       &       &   K      &   Hz                   &         \\
\hline
 
   73.2860      & $^2\Pi_{3/2},J=3/2, \frac{5}{2}(-)$   & $^2\Pi_{3/2},J=3/2, \frac{3}{2}(+)$  & 51    & 50   & 179.72616 & 1.108$\times$10$^{-17}$ &         \\
   74.9310      & $^2\Pi_{3/2},J=3/2, \frac{5}{2}(+)$   & $^2\Pi_{3/2},J=3/2, \frac{3}{2}(-)$  & 52    & 49   & 179.72621 & 1.184$\times$10$^{-17}$ &         \\
   80.1814      & $^2\Pi_{1/2},J=7/2, \frac{9}{2}(+)$   & $^2\Pi_{1/2},J=7/2, \frac{7}{2}(+)$  & 22    & 21   &  36.12975 & 8.253$\times$10$^{-20}$ & $\Delta P=0$ \\
   88.1811      & $^2\Pi_{1/2},J=5/2, \frac{7}{2}(-)$   & $^2\Pi_{1/2},J=5/2, \frac{5}{2}(-)$  & 16    & 15   &  19.28224 & 8.734$\times$10$^{-20}$ & $\Delta P=0$ \\
  107.3682      & $^2\Pi_{1/2},J=3/2, \frac{5}{2}(+)$   & $^2\Pi_{1/2},J=3/2, \frac{3}{2}(+)$  & 10    &  9   &   7.24584 & 7.995$\times$10$^{-20}$ & $\Delta P=0$ \\
  205.9510      & $^2\Pi_{1/2},J=1/2, \frac{1}{2}(-)$   & $^2\Pi_{1/2},J=1/2, \frac{1}{2}(+)$  &  3    &  2   &   0.01079 & 9.518$\times$10$^{-17}$ &         \\
  225.9357      & $^2\Pi_{1/2},J=1/2, \frac{1}{2}(-)$   & $^2\Pi_{1/2},J=1/2, \frac{3}{2}(+)$  &  3    &  1   &   0.01079 & 1.004$\times$10$^{-15}$ &         \\
  411.2056      & $^2\Pi_{1/2},J=1/2, \frac{3}{2}(-)$   & $^2\Pi_{1/2},J=1/2, \frac{1}{2}(+)$  &  4    &  2   &   0.02064 & 3.021$\times$10$^{-15}$ &         \\
  431.1905      & $^2\Pi_{1/2},J=1/2, \frac{3}{2}(-)$   & $^2\Pi_{1/2},J=1/2, \frac{3}{2}(+)$  &  4    &  1   &   0.02064 & 4.347$\times$10$^{-15}$ &         \\
\hline
\end{tabular}%
}
Frequencies are taken from the Leiden database. An entry $\Delta P = 0$ in the final column indicates a parity-conserving transition. Transitions of this type are included in the table only if they are in the observable frequency range of the {\it MWA}.
\end{table*}

\section{Observations}
The Murchison Widefield Array (MWA;\citealt{Tingay_2013,Wayth_2018}) consists of 256 tiles, of 16 dipoles each, arranged on the Murchison Radio-astronomy Observatory in Western Australia. The Phase-I array consisted of 128 dipole aperture tiles spread over 3\,km with a compact core. The Phase-II expanded array adds an additional 128 tiles, 56 of which are used to extend the baselines to 5.5 km.  During these observations only 91 tiles were online during building and commissioning of the Phase-II array with baselines between 1.5 and 5.5\,km. The dipoles themselves do not move and the mechanism to observe in the direction of a given source is to electronically apply delays. Whenever the control software determines the source is outside the sensitive region of the primary beam, the delays are reset at quantised values.  This means that each 5-minute observation samples a slightly different patch of sky. 

\subsection{Data Processing}
Each observation is calibrated using a 2-minute observation of Hydra A (a radio galaxy with a flux density of 243\,Jy at 160\,MHz (\citealt{Kuehr_1981}) from the beginning of each night. The bandpass and phase solutions are further refined using self-calibration of the calibrator field prior to applying the corrections to the observations of the target field. Each of the 5-minute observations of the target field are processed by first imaging each of the polyphase filter bank coarse channels (1.28\,MHz $\times$ 24) to check the data quality of the set of observations.  Of the 9 nights of observation (for a total of 30\,hours) between 05 January 2018 and 23 January 2018, four nights were not used, as they were impacted by severe radio frequency interference (RFI). Normally the RFI environment of the Murchison Radio-astronomy Observatory is clean but occasionally there are nights of periodic strong interference \citep{OffriingaRFI,Sokolowski_17}.  

For each of the remaining nights of observation, 100 of the 10\,kHz fine frequency channels per 1.28\,MHz coarse channels are imaged (for a total of 2400 channels) across the 30.72 MHz bandwidth. This frequency resolution corresponds to a velocity resolution of 24--30\,km\,s$^{-1}$ for objects within our own Galaxy. The reason for only imaging 78\% of the band is to avoid fine frequency channels known to be affected by aliasing due to the filter bank used to channelise the data.

The Phase-II configuration of the MWA used in these observations removed the compact core and had shortest baselines of 1.5\,km.  In order to obtain as much sensitivity to the diffuse emission as possible, all the images were created using Briggs weighting of ``0.5'' closer to natural weighting than uniform weighting.  Due to the large volume of data, it was not practical to process them using different weightings for this survey, even though the angular resolution was slightly compromised.  

Upon completion of the imaging of each fine channel for each observation, the channel images are combined into a 3-dimensional data cube using software written in {\sc Python}.  All of the cubes for all of the observations are then integrated together using the {\sc miriad} \citep{Sault} program {\sc imcomb}.  For greater detail on the data processing see \cite{Tremblay_2018}.

Free electrons in the Earth's atmosphere can create spatially varying refraction and propagation delays that are significant at low radio frequencies ($\ll$1\,GHz).  For each 5-minute observation, the estimated source positions are compared to the Molonglo Reference Catalogue (MRC; \citealt{Large_1981}) to correct for shifts in apparent positions.  After correction, the systematic spatial error in a fully integrated continuum image was --10$\pm$26\,arc\,seconds in right ascension and $+$6$\pm$11\,arc\,seconds in declination.  Both of these values are smaller than the synthesised beam of 1.03\,arc\,minute.  The ratio of point source flux density and peak intensity is 1.05.  This suggests that the ionospheric distortions are corrected well in these observations.

\subsection{Continuum subtraction and flagging}\label{sec:cont_flag}

The continuum signal in the cubes is formed from a combination of diffuse Galactic structure, including the Vela and Puppis supernova remnants, and hundreds of background extragalactic point sources, as well as the non-deconvolved sidelobes of structures both inside and outside the field-of-view. It varies slowly and smoothly along the frequency axis, except at the edges of the coarse channels, which are flagged.
To calculate the continuum to be subtracted from the cubes, we calculate a moving boxcar median along the spectral axis, across a range of 200\,kHz (20 fine channels). For channels which are within 100\,kHz of a coarse channel edge, we use the median value of the channel exactly 100\,kHz from the coarse channel edge. The resulting continuum cube is subtracted from the original cube to produce a noise-like cube which can be searched for spectral lines.

At this stage, channels with RFI contamination are easily identified by their high RMS. We calculate the mean RMS along the spectral axis and flag any channel which has noise 20\% higher than the mean value, repeating this process three times. These flags are applied to the \textit{original} data cube, and then the continuum subtraction is performed again. This ensures that RFI does not contaminate the continuum estimate. The first coarse channel ($\sim$99.2--100.5\,MHz) was entirely flagged for RFI and not used for further analysis.

\subsection{Survey Statistics}\label{sec:noise}

\begin{figure}
\centering
\includegraphics[width=0.48\textwidth]{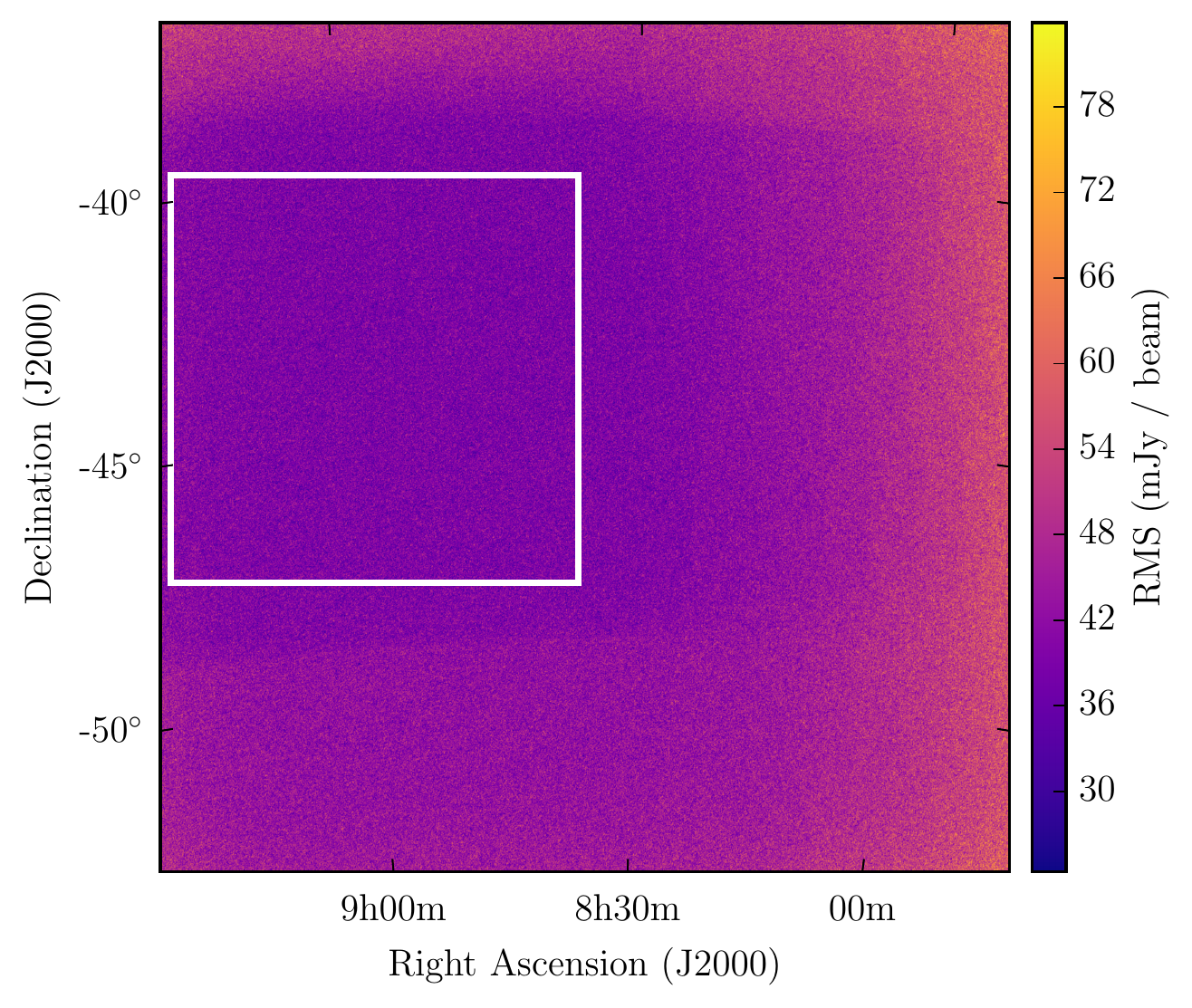}
\caption{An image of the RMS along the spectral axis of the flagged and continuum-subtracted 107.02--108.02-MHz cube. The white square indicates the region selected for subsequent noise analysis in Section~\ref{sec:noise}.}
\label{fig:rms_image}
\end{figure}

To search these data for lines, i.e. determine what signals are significant, we must first examine the properties of the noise in the data. Fig.~\ref{fig:rms_image} shows an image of the spatial distribution of the RMS noise, for the flagged and continuum-subtracted 107.02--108.02-MHz cube; the other channels look similar. The sensitivity of the stacked cube varies by over a factor of two over the region, due to the combination of many different fields-of-view with different primary beam sensitivities. There are no visible artefacts from the continuum subtraction or RFI-flagging. To examine the noise in more detail, we select a region of roughly constant noise, highlighted by a white box in Fig.~\ref{fig:rms_image}.

There are 24~coarse channel cubes, each comprised of $100\times10$\,kHz channels, and subtending $2500\times2500$~pixels. Since interferometers produce images which are correlated on the scale of the synthesised beam, to obtain the number of independent samples, we must divide by the synthesised beam volume $\frac{\pi a b}{4 \ln 2}$, where $a$ and $b$ are the major and minor full-widths-at-half-maxima of the synthesised beam: 165$''$ and 92$''$, respectively. Before flagging, there are therefore about 700\,million independent samples to search for lines. The flagging process described in Section~\ref{sec:cont_flag} removes 201 of the 2,400 initial channels, leaving 641\,million independent samples.

\begin{figure}
\centering
\includegraphics[width=0.48\textwidth]{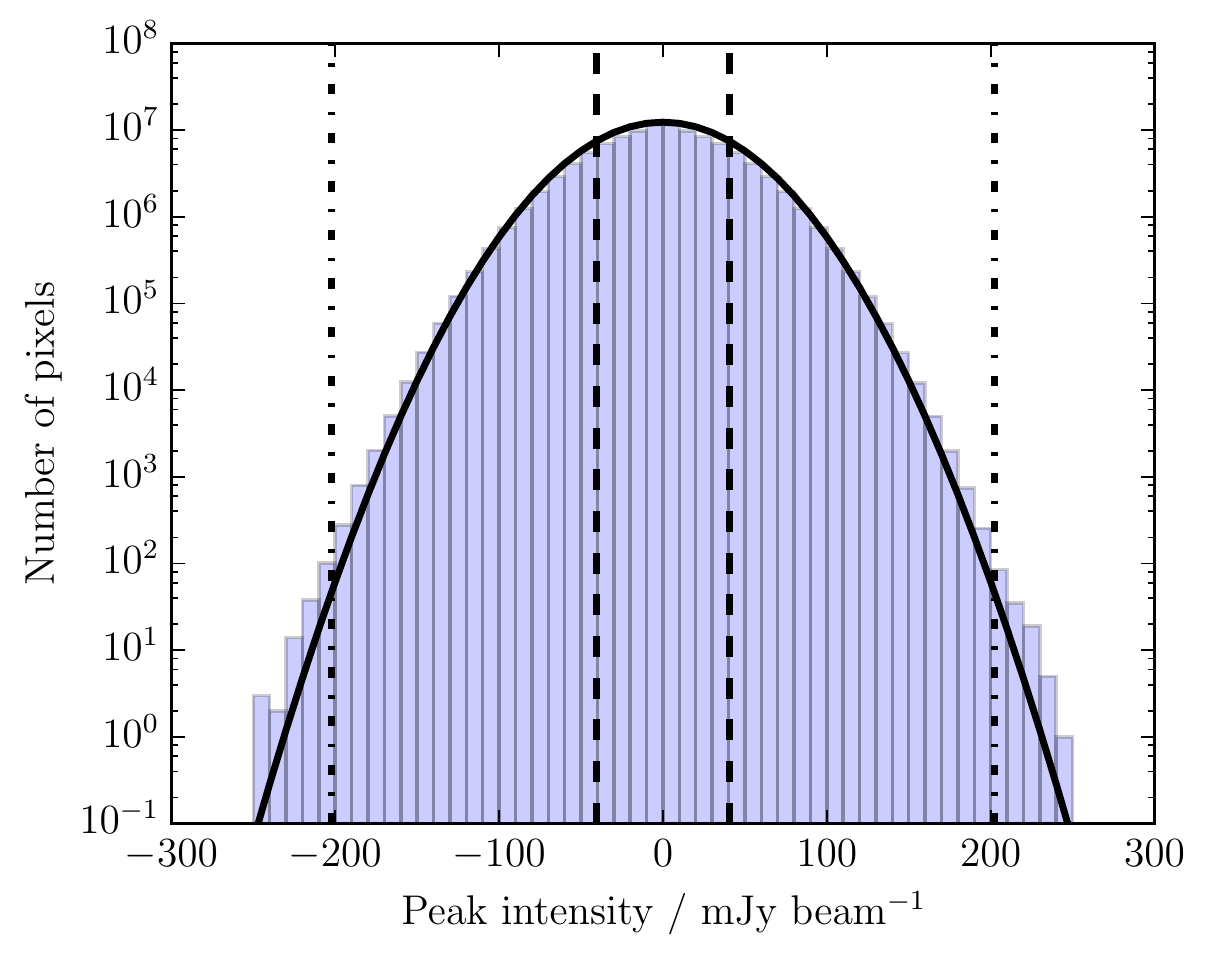}
\caption{A histogram of image pixels of the flagged and continuum-subtracted 107.02--108.02-MHz cube, for the white boxed region indicated in Fig.~\ref{fig:rms_image}. The standard deviation $\sigma$ of the distribution is 40.5\,mJy\,beam$^{-1}$, indicated by vertical black dashed lines. $5\sigma$ is shown by vertical dash-dotted lines. The black curve shows a Gaussian function (overlaid, not fit) with identical $\sigma$ to the data.}
\label{fig:rms_hist}
\end{figure}

Fig.~\ref{fig:rms_hist} shows a histogram of the pixel values in the white boxed region selected in Fig.~\ref{fig:rms_image}, compared to a Gaussian distribution of the same standard deviation. While the central part of the histogram resembles a Gaussian, the tails contain pixels with values in excess of those predicted by Gaussian statistics. A similar analysis of all the cubes shows that spurious values of up to $\pm 7 \sigma$ can appear in these cubes. This is not unexpected, as most interferometric data has non-Gaussian noise, and it is only the long integration time and the dense $(u,v)$-coverage of the MWA \citep{Wayth_PhaseII} that makes the distribution so close to Gaussian. 

\subsection{Source-finding}

Each of the 2400 continuum-subtracted fine-channel (10\,kHz) images are independently searched using the source-finding software {\sc Aegean} \citep{Hancock-2018}. This is done using the function ``slice", to set which channel in the cube is searched, and setting a ``seed clip" value of 5, in order to search the image for pixels with a peak intensity value greater than 6\,\,$\sigma$ (where $\sigma$ is set from an input RMS image such as that shown in Fig.~\ref{fig:rms_image}).  This source-finding threshold, based on the results of Section~\ref{sec:noise}, has the goal of detecting all signals $>$7\,$\sigma$\footnote{Some 7-$\sigma$ sources will be lost with a seedclip of $7\sigma$ since they may not be pixel-centred.}.

\section{Survey Strategy}
\begin{figure*}
\centering
\includegraphics[width=0.98\textwidth]{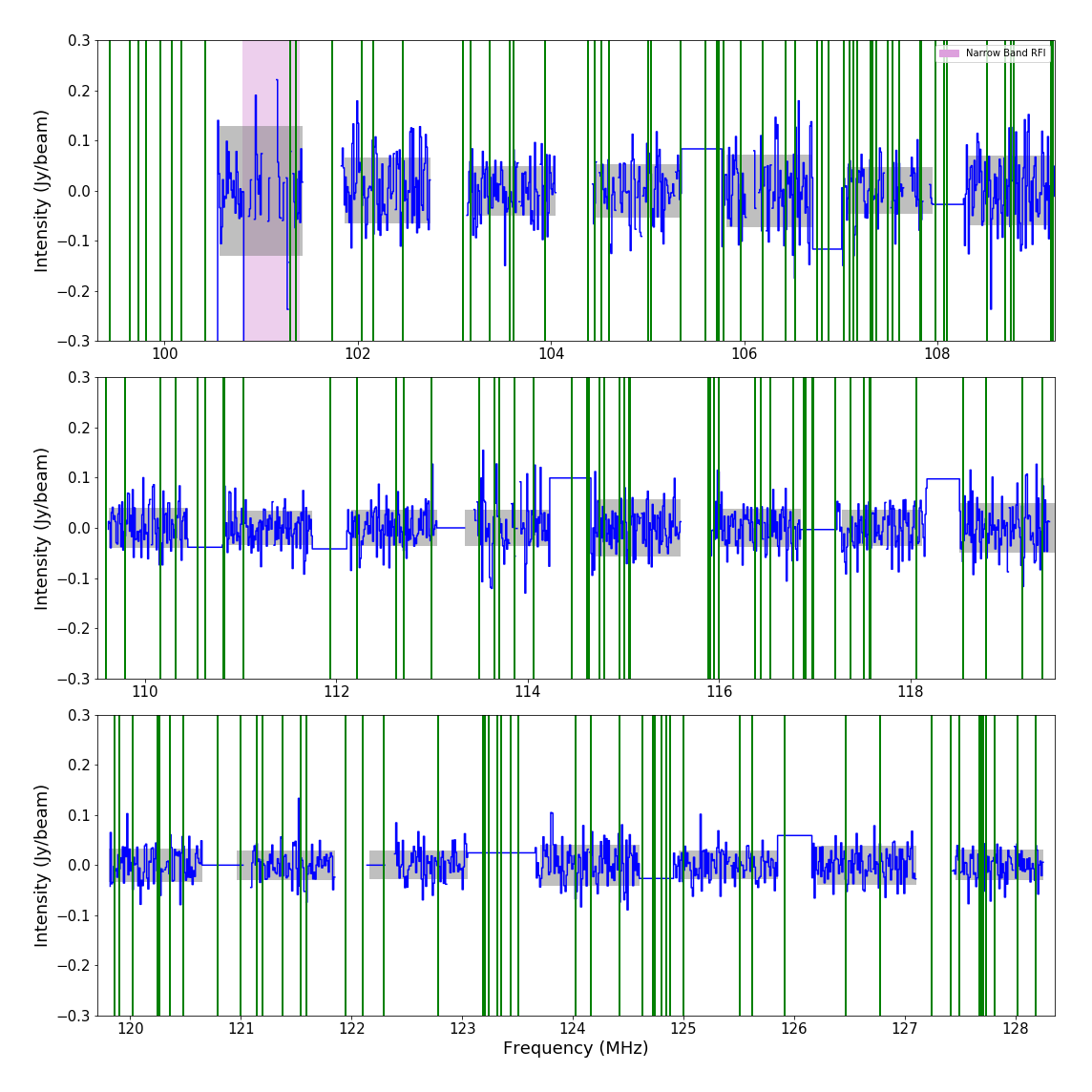}
\caption{The full spectrum of the MWA band towards RA=07:49:30 and Dec=--51:53:27 (star identified in $Gaia$ as TYC 8146-1566-1).  The green vertical lines represent the 192 known molecular transitions in the band.  The pink highlighted region is the region of residual narrow band RFI and the blank spaces are flagged channels.  The grey shaded region represents the one sigma RMS for each of the data cubes.  One data cube ($\sim$99\,MHz) was fully flagged due to significant RFI and the second cube ($\sim$100\,MHz) is flagged at this position. }
\label{spectra}
\end{figure*}

A blind spectral line survey of $\sim$200\,square\,degrees was completed with the MWA across the bandwidth of 98--129\,MHz.  In this band there are 196 known molecular transitions with upper energies less than 300\,K.  However, there has been little assessment of which of these lines are likely to be found in astrophysical environments.  We use the limit of 300\,K as above this limit it is expected that the transitions would be unlikely to be detectable in astrophysical environments, as the number of molecules within these kinetic temperatures is likely small.  Fig. \ref{spectra} shows the full spectrum of the band with vertical lines showing the population density of the known transitions. Of the known transitions, most of the rest frequencies are calculated theoretically and there are often only one or two lines for each molecular species. To identify potential peaks in the survey, we used the following databases: Cologne Database for Molecular Spectroscopy (CDMS; \citealt{Muller_2001}); Spectral Line Atlas of Interstellar Molecules (SLAIM;  Splatalogue\footnote{https://www.cv.nrao.edu/php/splat/index.php}); Jet Propulsion
Laboratory (JPL; \citealt{Pickett_1998}); and Top Model \citep{Carvajal_TopModel}.

\section{Results}
At a significance level of 7\,$\sigma$ we did not detect any signals in emission.  This is a limit of approximately 0.21--0.35\,Jy\,beam$^{-1}$ (depending on the specific region) and is a flux density limit that matches (or is above) our previous tentative detections in our surveys toward the Galactic Centre and Orion.  This translates to a single-channel upper limit of 26,760\,K\,km\,s$^{-1}$ for the 107\,MHz transition of nitric oxide.  This would also mean the column densities would be higher than the upper limit set in \cite{Tremblay_2017}. It is therefore unsurprising that we didn't detect anything at this level. In this survey we use the early science data from Phase II, whilst the previous survey was done with Phase I in a stable state after years of operation. Also, with the extended baselines we are more sensitive to different emission (dense gas versus diffuse). 

It is possible that some features may be in absorption against strong continuum.  To assess if we detect any absorption features, we created spectra toward five of the brightest continuum sources in the field, consisting of the Puppis Supernova Remnant, star LBS72 Star E, the radio continuum objects CUL 0836-443 \citep{1995AuJPh..48..143S} and PMN J0820-4736 \citep{2012MNRAS.422.1527M} and the H{\sc ii} region RCW 38. We found the continuum emission brightness varied for these sources from 0.4 to 7\,Jy\,beam$^{-1}$ with a median optical depth limit of 0.22$\pm$0.13. Within the non-continuum subtracted data cubes, no signals were detected over a 7\,$\sigma$ limit in absorption\footnote {We note that the higher noise is a result of residual bandpass structure around bright continuum sources which is normally corrected for in the continuum subtraction}.

\section{Discussion}
It is known that low-frequency molecular lines are weak emitters (e.g. \citealt{Codella_2015}), which makes this type of experiment, in particular with a 24\,km\,s$^{-1}$ velocity resolution, a difficult one. However, low frequency transitions are, in general, relatively easy to invert \citep[e.g.][]{elitzur1992}, making the MWA frequency range a potentially interesting hunting ground for new masers.  

One of the largest molecules of interest by the astrochemical community is the amino acid, glycine (NH$_{2}$CH$_{2}$COOH).  \cite{Codella_2015} predicts that to observe glycine at centimetre wavelengths would require over 1000 hours of observation to obtain a three-sigma detection.  This is a considerable effort in data processing, as well as the observing time commitment. Our survey represents a start in understanding some of these data challenges as we approach the era of the SKA.

In this survey, the $\sim$30\,hours of observations taken in five-minute snap-shots, represents around 360 individual raw visibility data sets.  Each data set was calibrated and the continuum image was created for each of the 24 coarse frequency bands.  For any observation or night of observations that were affected by imaging artefacts or severe RFI, no further processing was completed.  For the remaining observations, each of the 2400 fine frequency channels were independently imaged. This created more than 443,000 continuum images and used over 350,000 CPU hours on the `Magnus' computing cluster at the Pawsey SuperComputing Centre and over 300\,TB of disk space.  This represents significant computing resources for a single project. These values do not include the resources required to then search and create visual representations for scientific analysis. 

Additional challenges regarding spectral line data processing with SKA precursors include making compromises on data quality, field-of-view processed, or number of spectral channels processed based on amount of RAM, processing time, and read/write speeds on the computer.  This limits the amount of the data taken by the telescopes which is available for actual science and impacts blind surveys such as this work.  There are also significant challenges regarding visualising data cubes that are hundreds of GB in size. This is being addressed by the community with new tools like {\sc carta} \citep{angus_comrie_2020_3746095}. The requirements for this style of data processing are not likely to get smaller with the next generation of telescopes; it is therefore important to understand the data processing challenges with surveys such as this.

\section{Conclusion}
This work had two goals: to complete the deepest survey with the MWA than previously attempted; and to increase our knowledge of emission lines of $^{14}$NO and $^{15}$NO. We completed a deep spectral line survey toward the Vela region, but found no signals at a peak intensity limit of 0.21\,Jy\,beam$^{-1}$.  This survey is the third in a series completed with the MWA, but the first using the new extended baselines, for an improved resolution of 1\,arc\,minute (versus the original 3\,arc\,minutes).  We have also found that the noise no longer decreases as a function of the square root of time, suggesting that other improvements in the data processing are required to obtain a better result.  It is likely the noise is limited to the deconvolution of individual snapshot images of the fine frequency channels.

The molecular modelling presented in Section 2 for Nitric Oxide gives us precise frequency targets to search for with the MWA and other SKA precursor instruments, like the Parkes 64\,m Telescope  and the Australia Square Kilometre Array Pathfinder .  Many of the new lines are within the frequency range of the new Parkes ultra-wide band receiver (704--4032\,MHz, \citealt{Hobbs_UWL}), making it an interesting choice for future simultaneous searches of NO, CH and OH. 

Overall, we present this work as an important step in understanding the data processing challenges we will face with the Square Kilometre Array and places solid limits on what can be expected for these low-frequency surveys in the future.

\section{acknowledgements}
\subsection{Personnel}
We thank both of the anonymous reviewers for their comments that improved the quality of this manuscript. NHW is supported by an Australian Research Council Future Fellowship (project number FT190100231) funded by the Australian Government.  
\subsection {Facilities}
This scientific work makes use of the Murchison Radio-astronomy Observatory, operated by CSIRO. We acknowledge the Wajarri Yamatji people as the traditional owners of the Observatory site. Support for the operation of the MWA is provided by the Australian Government (NCRIS), under a contract to Curtin University administered by Astronomy Australia Limited.  Establishment of ASKAP, the Murchison Radio-astronomy Observatory and the Pawsey Supercomputing Centre are initiatives of the Australian Government, with support from the Government of Western Australia and the Science and Industry Endowment Fund. 
\subsection{Computer Services}
We acknowledge the Pawsey Supercomputing Centre which is supported by the Western Australian and Australian Governments. Access to Pawsey Data Storage Services is governed by a Data Storage and Management Policy (DSMP). ASVO has received funding from the Australian Commonwealth Government through the National eResearch Collaboration Tools and Resources (NeCTAR) Project, the Australian National Data Service (ANDS), and the National Collaborative Research Infrastructure Strategy. This research has made use of NASA’s Astrophysics Data System Bibliographic Services.  The kinetic data we used have been downloaded from the online database KIDA (Wakelam et al. 2012, http://kida.obs.u-bordeaux1.fr).
\subsection{Software}
The following software was used in the creation of the data cubes:
\begin{itemize}
    \item {\sc aoflagger} and {\sc cotter} -- \cite{OffriingaRFI}
    \item {\sc wsclean} -- \cite{offringa-wsclean-2014,offringa-wsclean-2017}
    \item {\sc Aegean} -- \cite{Hancock_2018_Aegean}
    \item {\sc miriad} -- \cite{Sault}
    \item {\sc topcat} -- \cite{Topcat}
    \item{NumPy v1.11.3 \citep{NumPy}, AstroPy v2.0.6 \citep{Astropy}, SciPy v0.17.0 \citep{SciPy}, Matplotlib v1.5.3 \citep{Matplotlib}}
    \item {\sc CARTA} -- \cite{angus_comrie_2020_3746095}
\end{itemize}

\bibliographystyle{aasjournal}
\bibliography{Vela_Survey}

\end{document}